\documentclass[12pt]{article}

\usepackage{mathrsfs,amsbsy,amssymb,latexsym,amsfonts,amsmath,cite}

\global\arraycolsep=3pt
\newcommand\CL{{\mathcal L}}

\begin{document}

\begin{flushright}
        CERN-PH-TH/2008-035\\ 
	YITP-07-99\\
	OIQP-07-20
\end{flushright}

\begin{center}
{\Large Test of Effect from Future in Large Hadron Collider;\\A Proposal}

\vspace{10mm}

{\large
Holger B. Nielsen}
{\it
\footnote{
On leave of absence to CERN, Geneva from 1 Aug. 2007 to 31 March 2008.}\\
The Niels Bohr Institute,
University of Copenhagen,
\\
Copenhagen $\phi$, DK2100, Denmark
}
\\
and
\\
{\large
Masao Ninomiya}
{\it
\footnote{
Also working at Okayama Institute for Quantum
Physics, Kyoyama 1, Okayama 700-0015, Japan.}\\ 
Yukawa Institute for Theoretical Physics,\\
Kyoto University, Kyoto 606-8502, Japan}\\

\end{center}

\begin{abstract}
We have previously proposed the idea of performing a card-drawing experiment of which 
the outcome potentially decides whether the Large Hadron Collider (LHC) 
should be closed or not. The purpose is to test theoretical models such as 
our own model that have an action with an imaginary part that has a 
similar form to the real part. The imaginary part affects the 
initial conditions not only in the past but even from the future. It was 
speculated that all the accelerators producing large amounts of Higgs particles 
such as the Superconducting Super Collider (SSC) would mean that the initial 
conditions must have been arranged so as not to allow these accelerators 
to work. If such effects existed, we could perhaps cause a very 
clear-cut ``miracle" by having the effect of a drawn card to be the  
closure of the LHC. Here we shall, however, argue that the closure of an accelerator
is hardly needed to demonstrate such an effect
and seek to calculate how one could perform a verification experiment for the 
proposed type of effect from the future 
in the statistically least disturbing and least harmful way.

We shall also discuss how to extract the maximum amount of 
information about such as effect or 
model in the unlikely case that a card preventing the running of the LHC or the 
Tevatron is drawn, by estimating  
the relative importance 
of high beam energy or high luminosity for the purpose of our effect.
\end{abstract}

\section{Introduction}

~~~~Each time an accelerator is used to investigate
a hitherto uninvestigated regime 
such as collision energy or luminosity, there is, a priori, a 
chance of finding new effects that, in principle, could mean that 
a well-established principle could be violated,
in lower-energy physics or in daily life. 
The present paper 
is one of a series of articles \cite{search,3,4,5,14} discussing how one might 
use the LHC and perhaps 
the Tevatron to search for effects violating the following
well-established principle while the future is very much influenced by the past, 
the future does not influence the past. Perhaps we can more precisely state this 
principle, which we propose to test at the new LHC accelerator, as follows: 
While we find that there is a lot of structure from the past that exists today  
in its present state -- at the level of pure physics --
simple structures existing in the future, so to speak, 
do not appear to 
prearrange the past so 
that they are \cite{r1,7,6,8,9,10,11 }. If there really 
were such prearrangements organizing simple things to exist 
in the future we could say that it would be a model for an initial state 
with a built-in arrangement for the future, which is what our model is.
However,
models or theories for the initial state such as Hartle and Hawking's 
no-boundary model \cite{2} are not normally of this 
type, but rather lead to a simple starting state corresponding to the fact that we 
normally do not see things being arranged for the future in the fundamental 
laws and thus find no backward causation \cite{r1}. 
However, 
we sometimes see 
that this type
of prearrangement occurs, but we manage to explain it away. For example, 
we may see lot of people gathering for a concert. At first it 
appears that we have a simple structure in the future, namely, many people 
sitting in a specific place, such as the concert hall, causing a prearrangement in 
the past.

Normally we do not accept the phenomenon of people gathering for a concert 
as an effect of some mysterious fundamental physical law seeking to collect the 
people at the concert hall, and thus arranging the motion of these 
people shortly before the concert to be directed 
towards the hall. In our previous model  
\cite{search,3,4,5,6,14}, which even we do not claim to be 
relevant to 
the concert hall example, such an explanation based on 
a fundamental physics model 
could have sounded plausible. In our model, we have a quantity $S_I$, 
which is the imaginary part of an action in the sense that it is substituted 
into a certain Feynman path integral,  as is 
the real part of the 
action, except for a factor $i$. In fact, we let  the action $S$ be 
complex, and its imaginary part $S_I$, as for the real part $S_R$,
be an integral 
over time, $S_I=\int L_Idt$. 
Thus 
$S=S_R+iS_I$. 
Roughly speaking, the way that the world develops is  
to make $S_I[path]$ almost minimal (so that the probability weight obtained from 
the Feynman path integral $e^{-2S_I}$ is as large as possible). Thus, a 
tempting ``explanation" for the gathering of the people would be that  
many people gathering for a concert provides a considerable negative 
contribution to the imaginary part $L_I$ of the Lagrangian during the concert, 
and thereby, a negative contribution to the imaginary action $S_I$. Thus, the 
solutions to the equations for such a gathering before a concert would have 
an increased probability of $e^{-2S_I}$, and we would have an explanation for the 
phenomenon of the people gathering for the concert. If we did not have an 
alternative -- and we think better -- explanation, then we might have to take 
such gatherings of people for concerts as  evidence for our type of model 
with an effect from the future; we would, for example, 
conclude that the gathering of people 
occurred in order to minimize  ``an imaginary action" $S_I$.

An alternative and better explanation that does not require any fundamental 
physical influence from the future is as follows: 
The participants in the concert and 
their behaviors are indeed, in the classical and naive approximation,
completely 
determined from the initial state of the universe at the moment of the Big Bang,
an initial state in which the concert was not planned. Later on, however, some 
organizers -- possibly the musicians themselves -- used their phantasy to model the 
future by means of calendars, etc., and they issued an announcement. We can 
use this announcement as the true explanation of the gathering of the 
listeners to the concert. The gathering at the concert was due to 
some practical knowledge of the equation of motion allowing the 
possibility to organize events entirely on the basis of the 
equation of motion and using the fact that the 
properties of the initial conditions are, in some respects, very well 
organized (low entropy, sufficient food and gasoline resources). However, there was no 
effect from the future, only from the phantasies about the  
future implemented in memories, which are true physical objects of course, such 
as the biological memories of the announcement and so on.

Even more difficult examples can be used to explain 
the fact that our actions are
not preorganized 
``by God", which may here be roughly identified with fundamental physical
influences from the future, such as the biological development of extremely 
useful organs. Has the development of legs, say, really got nothing to do with 
the fact that 
they can later be used for walking and running? Darwin 
and Walles produced a convincing explanation  
for the development 
of legs without the need for any fundamental influence from the future on the past.

If, as would be said prior to Darwin's time, it were God's plan 
(analogous to the concert organizer's plan) to make legs, this would come very 
close to the fundamental physics model, provided the following two assumptions
were satisfied: 
\begin{enumerate}
\item[1)] That this God is not limited or has His memory limited by
physical degrees of freedom in contrast to the brains of the 
concert organizers.
\item[2)] This God is all-knowing, which means that He has access to the 
future and does not need a phantasy or simulation to create a model of 
it.
\end{enumerate}

In the earlier works\cite{3,4,5} we attempted to find 
various reasons why this effect 
from the future might
to be suppressed. For instance this effect is
definitely suppressed for particles whose eigentimes are trivial 
in some sense. 
From the Lorentz invariance, the contribution of an action, 
which may be to the real part $S_R$ or 
the imaginary part $S_I$ involving from the passage of a particle from one point to 
another point must be proportional to the eigentime of the passage (i.e., the 
time the passage would take according to a standard clock located at the particle).

Two examples that dominate the physics of daily life ensure at least one 
source of strong suppression of the effect from the future: 
1) Massless particles such as the photon have always zero eigentimes, thus for 
photons, the effect is strongly suppressed or killed. 2) For nonrelativistic 
particles, the eigentime is equal to the reference frame time, and thus the eigentime 
is trivial unless the particle is produced and/or destroyed. If a 
particle such as an electron is conserved the eigentime becomes trivial and 
there is little chance to see our effect at the lowest order with electrons.

Actually, since there is a factor of $\frac{1}{\hbar}$ in front of the 
action, one initially expects the effects of $S_I$ to be so large that we 
need a large amount of suppression to prevent our model from being in
immediate disagreement with the experiment.

We shall discuss some suppression mechanisms in section 3. We shall also 
discuss reasons why it may be likely that 
the effects of Higgs 
particles are much greater than those of already observed particles. 
The number of Higgs particles is not preserved so that even a 
non relativistic Higgs particle may contribute $S_I$, 
but we do, of course, expect 
somewhat relativistic Higgs particles to be produced 
with velocities of the order of magnitude of that of light, but typically
not extremely relativistic,
thus there is no reason from 
the two above mentioned mechanisms that the effect of a Higgs particle should 
be suppressed.

In contrast, as we shall see, there is a reason why even if  the whole 
effect of $S_I$ is generally strongly suppressed, a counteracting 
factor could be caused for the case of the Higgs particle.

Normally it would be reasonable to assume that in the real and imaginary parts, the
multipliers of the same field combination, say  $\lambda_R$ in 
$\frac{\lambda_R}{4}|\phi |^4$ in $S_R$ and  $\lambda_I$ in 
$\frac{\lambda_I}{4}|\phi |^4$ in $S_I$, should be of the same order of 
magnitude, so that the complex phases for the couplings, such as  
$\lambda_R+i\lambda_I$, should be of the order of unity.

There is, however, one case in which the problems connected with the 
hierarchy problem make this assumption unlikely to be true: 
Because of the hierarchy problem it is difficult to avoid considerable 
fine tuning of 
the Higgs mass, since its quadratically divergent contributions with a cutoff 
at the Planck scale $M_{Pl}$ would shift it considerably. 
We may only need this fine tuning for the squared term in the 
real part of the complex mass 
$m^2|\phi |^2=\left(m_R^2+im_I^2\right)|\phi |^2$ in the 
Lagrangian density. 
Whether or not only the real part $m_R^2$ of 
the square of the mass is tuned or whether also the imaginary part $m_I^2$
is also tuned 
may depend on which of the various models is used to attempt to solve 
the hierarchy problem, and how such a model is implemented together 
with our model of the imaginary part of the action.

For instance, one of us has constructed a long argument
-- using a  
bound state of six top and six antitop quarks -- that under the assumption of 
several degenerate vacua (=MPP)[\cite{14,7,13}], which in turn 
follows \cite{14} from the model in the present article, we obtain 
a very small Higgs mass with its order of magnitude agreeing with 
the weak scale. In this model, which ``solves the hierarchy problem", 
it is clearly the 
real part of the square of the mass, i.e., $m_R^2$, that gets fine tuned, 
and there 
would be insufficiently many equations of MPP stating that several vacua 
have essentially zero (effective) cosmological constants allowing the fine 
tuning of more 
than just this real part $m_R^2$. In this model, the 
argument would thus be that 
the hierarchy problem would remain unsolved for the imaginary part of the 
coefficient of the 
mass-square term for the Higgs field $m_I^2$, but 
it is unknown whether 
this imaginary part $m_I^2$ should be fine tuned to be small. A priori, 
it may be difficult for any model to obtain a small real part $m_R^2$ 
of the weak scale; thus it is highly possible that also in other models, 
only the real part $m_R^2$ is tuned and not the imaginary part. In such 
cases, the imaginary part $m_I^2$ of the Higgs mass square could, a priori, 
remain untuned and be of, the order of some fundamental scale, 
such as the Planck scale or a unified 
scale. This would mean that the imaginary part 
$m_I^2$ may be much larger than the real part, 
\begin{align}
	m_I^2\gg m_R^2 
~,
\end{align}
and thus, the assumption that all the ratios of the real to the imaginary part 
for the various coefficients in the Lagrangian should be of order unity would 
not be expected to be true for the case of the mass-square coefficient. 
Conversely, unless the hierarchy problem solution is valid for both real and 
imaginary parts we could have 
\begin{align}
	\frac{m_I^2}{m_R^2}\simeq \frac{M_{PL}^2}{M_{weak}^2}\simeq 
	\left(\frac{10^{19}\text{ GeV}}{100\text{ GeV}}\right)^2\simeq 10^{34}.
\end{align}
This would mean that by estimating the effect of $S_I$ in an analogous way 
to that for particles with real and imaginary parts of their 
coupling coefficients being of the same order or given by some general 
suppression factor, we could potentially 
under estimate the effects of the Higgs particle 
by a factor of $10^{34}$.

In the light of this estimation for the relative importance of the effect
from the future ($\simeq$ our $S_I$ effect) on the Higgs particle 
relative to that on other particles, it is an obvious conclusion that one should 
search for this type of effect when new Higgs-particle-producing machines 
such as the LHC, the Tevatron, or the canceled SSC are planned.

If the production and existence of a Higgs particle for a small amount of time 
gave a 
negative contribution to $S_I$, which would enhance the probability density 
$\propto e^{-2S_I}$, one could wonder why the universe is not 
filled with Higgs particles. This may only be a weak argument, but 
it suggests that presumably the contribution from an existing Higgs 
particle to $S_I$ is positive if it is at all important. Now if the production 
and ``existence" of a Higgs particle indeed gave a positive contribution to $S_I$, 
whereby the probability $\propto e^{-2S_I}$ for developments in a world
containing large Higgs-particle-producing accelerators should be
decreased by the effect of their $S_I$ contribution, then 
the production and existence of 
such Higgs particles in greater amounts should be avoided somehow in the 
true history of the world. 
If an 
accelerator potentially existed that could generate a large number 
of Higgs particles and if the parameters were so that 
such an accelerator would
indeed give a large 
positive contribution, then such a machine should practically never be 
realized!

We consider this to be an interesting example and weak experimental evidence for our 
model because the great Higgs-particle-producing accelerator SSC\cite{16}, in 
spite of the tunnel being a quarter built, was canceled by  
Congress! Such a cancellation after a huge investment is already in itself an 
unusual event that should not happen too often. We might take this event 
as experimental evidence for our model in which an accelerator 
with the luminosity and beam energy of the SSC will not be built 
(because in our model, $S_I$ will become too large, i.e., less 
negative, if such an accelerator was built)\cite{16}.

Since the LHC has a performance approaching the SSC, it 
suggests that also the LHC may be in danger of 
being closed under mysterious circumstances.
In an introductory article to the present one 
\cite{search} we offered to demonstrate the mysterious effect of $S_I$
in our model on potentially closing the LHC by carrying out
a card-drawing game, or using a random number generator.

Under the assumption that our model is indeed correct, demonstrating a strong 
effect on the LHC by a card-drawing game such as its possible closure 
would serve a couple of purposes: 
\begin{enumerate}
\item[1)] Even though some unusual political or natural catastrophe causing the 
closure of the LHC would be strong evidence for the validity of 
a model of our type with an effect from the future, 
it would still be debatable whether the closure was not due to some 
other cause other than our $S_I$-effects.
However, if a card-drawing game or a quantum random number generator 
causes the closure of the LHC in 
spite of the fact that it was assigned a small probability of the order of, 
say $10^{-7}$, then the closure would appear to very clear evidence for our model. 
In other words, if our model was 
true we would obtain a very \underline{clear} evidence using such a 
card-drawing game or random number generator.
\item[2)] A drawn card or a random number causing a restriction on the LHC could be 
much milder than a closure caused by other means due to the 
effect of our model. The latter could, in addition, result in  the LHC 
machine being badly used, or cause other effects such as the total closure of CERN, 
a political crisis, or the loss of many human lives in the case 
of a natural catastrophe.

Thus, the cheapest way of closing the main part of the LHC may be 
to demonstrate the  effect via the card-drawing game.
\end{enumerate}

However, in spite of these benefits of performing the card-drawing experiment 
it would be a terrible waste if a card really did enforce the closure or a 
restriction on the LHC. It should occur with such a low probability under normal 
conditions that if our model were nonsense, then drawing a card requiring a strong 
restriction should mean that our type of theory was 
established solely on the basis of 
that ``miraculous" drawing. 
Such a drawing would have
the consolation that instead of finding supersymmetric partners 
or other novel phenomena 
at the LHC, one would see the influence from the 
future! That might indeed lead to even more spectacular new physics than one could 
otherwise hope for! 
Thus, the restriction of the LHC would not be so bad. Nevertheless, it 
is of high importance that one statistically 
minimizes the harm done by 
an experiment such as a card-drawing game. Also, one should 
allow several 
possible restrictions to be written on the cards that might be chosen 
so that several possible effects from the future 
may occur. Then one might, in principle, learn 
about the detailed properties of this effect
such as the number of Higgs particles needed 
to obtain an effect, or whether luminosity or beam energy matters the most
for the $S_I$?

It is the purpose of the present article to raise and discuss these questions 
of how to arrange a card-drawing game experiment to obtain 
maximum information and benefit and minimal loss and obtain statistically
minimal restrictions.

In section 2, we formulate some of the goals one 
would have to consider in planning a card-drawing game or random number 
experiment on restricting the LHC. In section 3 we give a 
simplified description of the optimal organization of the game and propose 
that the probability distribution in the game should be assigned 
on the basis of the 
maximum allowed size of some quantity consisting of luminosity, beam energy
and number of Higgs particles etc..

In section 4 we develop the theory of our imaginary action so as to 
obtain a method of estimating the mathematical form expected for the 
probability of obtaining a  
``miraculous" closedown of the LHC.

In section 5 we discuss possible rules for the card-drawing game 
and random quantum generator.  However, we still 
think that more discussion and calculation may be needed to develop the 
proposed example before actually drawing the cards.

Section 6 is devoted to further discussion and a conclusion.

\section{The goals, or what to optimize}

~~~~It is clear that the most important goal concerning the LHC is 
for it to operate in a 
way that delivers as many valuable and interesting results as possible 
while searching for one or more Higgs particles, strange bound states and 
supersymmetric partners.
In contrast, our model is extremely unlikely to be true, and thus, the 
investigation of our model should only be allowed to disturb the 
other investigations very marginally. The problem is that although the 
probability of disturbance 
by the investigation of our theory has been statistically 
evaluated to be very tiny, there is a risk that the selection of a very unlucky 
card could impose a significant restriction on the LHC, and thereby cause 
a very major disturbance.

\subsection{What to expect}

~~~~Before estimating the optimal strategy with respect to the card-drawing 
game and its rules, we wish to obtain a crude statistical impression of 
what to expect. 

The most likely event is that our model is simply wrong, and thus it 
is very unlikely that anything should happen to the LHC unless it
is caused 
by our card-drawing experiments. If, however, our model is
correct in principle, we must accept the unlucky fate of the SSC \cite{16} as 
experimental evidence and conclude that the amount of superhigh-energy 
physics at SSC, 
measured in some way by a combination of the luminosity 
and beam energy, seemingly sufficient to change the fate of the 
universe on a macroscopic scale. We do not at present know the parameters of 
our model, and even if we make order-of-magnitude guesses, 
there are several difficulties in estimating even order-of-magnitude 
suppression mechanisms. For instance, there may be competition between the 
arrangements of the events in our time to give a low $S_I$ with a similar 
arrangement for other times. 
Thus, even guessing the 
order-of-magnitude 
of fundamental couplings will still not give a safe estimate 
for the order-of-magnitude of the strength in practice. 
We are therefore left with a crude method of  
prior estimation by taking the probability of different 
``amounts of superhigh-energy collisions" (say some combination of beam 
energy and luminosity) needed to macroscopically change the fate of the 
universe to be of constant density in the logarithm of this measure of 
superhigh-energy collisions.

For simplicity, we consider that $\chi$ is, for example, the integrated 
luminosity 
for collisions with sufficiently high energy to produce Higgs particles, or take it 
simply to be the number of Higgs particles produced. Whether they are observed 
or not does not matter; it is the physically produced Higgs particles and the 
time they exist that matters.

We know in the case that our theory was correct, an upper limit for this 
amount of superhigh-energy collisions is needed to obtain an effect. 
Namely, we know that the SSC was canceled and that its potential amount of 
superhigh-energy collisions 
must have been above the amount needed. On the other hand, we also 
know that the Tevatron seemingly operates as expected
so that its 
amount of superhigh-energy collisions must be below the amount needed to cause 
fatal macroscopic changes.

In our prior estimation, we should thus calculate the probability for the 
amount $\chi$. 
This is needed to cause fatal effects in a machine, 
where $\chi$ is in the interval 
$\left[\chi -\frac{1}{2}d\chi ,\chi +\frac{1}{2}d\chi \right]$, as 
\begin{align}
	\text{Probability}\left(\left[\chi -\frac{1}{2}d\chi ,
	\chi +\frac{1}{2}d\chi \right]\right)=p(\chi )d\chi ,
\end{align}
and we assume 
\begin{align}
	p(\chi )=\frac{1}{\chi}
	-\frac{1}{\log \chi_{\text{SSC}}-\log \chi_{\text{Tevatron}}}
\end{align}
for $\chi_{\text{SSC}}\geq \chi \geq \chi_{\text{Tevatron}}$ and that
$p(\chi )=0$ outside this range.

Here we have respectively denoted the amounts of superhigh-energy collisions, say, 
the numbers of Higgs particles, produced as integrated luminosity in the SSC and 
the Tevatron by $\chi_{\text{SSC}}$ and $\chi_{\text{Tevatron}}$.

The SSC should have achieved a luminosity of $10^{33}$ cm$^{-2}$s$^{-1}$ 
and a beam energy $20$ TeV in each beam, while the Tevatron has achieved values of
$\sim 10^{32}$ cm$^{-2}$s$^{-1}$ and $1$ TeV.

The LHC should achieve a luminosity of $10^{34}$ cm$^{-2}$s$^{-1}$ and 
a beam energy of $7$ TeV in each beam.
\begin{center}
\begin{tabular}{|c||c|c|}\hline
 & Beam Energy & Luminosity \\ \hline \hline
Tevatron & $1$ TeV & $10^{32}$ cm$^{-2}$s$^{-1}$ \\ \hline
LHC & $7$ TeV & $10^{34}$ cm$^{-2}$s$^{-1}$ \\ \hline
SSC & $20$ TeV & $10^{33}$ cm$^{-2}$s$^{-1}$ \\ \hline
\end{tabular}
\end{center}

With respect to luminosity, the LHC is expected to be even stronger than the 
SSC; thus, if we apply our criterion, one would expect the LHC to be prone to even 
greater bad luck than the SSC.

Let us, however, illustrate the idea by using for the beam energy for $\chi$. 
Then  
\begin{align}
	& \log \chi_{\text{SSC}}=\log (20\text{ TeV}), \nonumber \\
	& \log \chi_{\text{Tevatron}}=\log (1\text{ TeV})
\end{align}
and 
\begin{align}
	\log \frac{\chi_{\text{SSC}}}{\chi_{\text{Tevatron}}}=\log 20\simeq 3.
\end{align}
Hence, 
\begin{align}
	p(\chi )=\frac{1}{\chi}\cdot \frac{1}{3}\qquad \text{for }
	\chi_{\text{Tevatron}}\leq \chi \leq \chi_{\text{SSC}}.
\end{align}
Now $\chi_{\text{LHC}}=\log (7\text{ TeV})\simeq 2+\log \text{TeV}$. 
Thus, the 
probability that the critical $\chi$ for closure is smaller than 
$\chi_{\text{LHC}}$ is $P(\chi <\chi_{\text{LHC}})=\frac{2}{3}$. 
This means that 
if our theory was correct and the beam energy was the relevant quantity, then 
the LHC would be stopped somehow with a probability of 
$\simeq \frac{2}{3}\simeq 66\%$.

\subsection{What would we like to know about our model, if it is correct?}

~~~~There are several information one would like 
to get concerning our model:
\begin{enumerate}
\item[1)] One would like to obtain an estimate of how strong the effect is, 
i.e., one would like to estimate at least the order-of-magnitude of the value of 
$\chi$ needed to disturb the fate of the universe macroscopically.
\item[2)] One would like to determine whether it is the beam 
energy or the luminosity that is most important for causing closing.
\item[3)] One would also like to determine which type of random numbers, 
quantum random numbers or more 
classically constructed ones, allow our $S_I$-effect to more easily
manipulate the past. 
One could even speculate whether one could construct a 
mathematically random number that should make it almost impossible to manipulate
such a 
\underline{physical} effect (from the future).
\end{enumerate}

In addition to all these wishes, to get questions answered
random number experiment that 
causes minimal harm to the optimal use of the LHC machine.

\subsection{How to  evaluate cost?}

~~~~Let us now discuss the above questions
about our model.

To obtain a convincing answer to question 1), of whether there is 
indeed an effect, as proposed, the probability of selecting 
a random number -- a card for instance -- that leads to restrictions should be so 
small that one could practically ignore the possibility that 
a restriction occurs
simply by chance. 
This suggests that one should let the a priori probability of a 
restriction, i.e., the number of card combinations corresponding to a 
restriction relative to the total number of 
combinations, be sufficiently small to correspond to getting by accident an 
experimental measurement five standard deviations 
away from the mean. 
That is to say, a 
crude number for the suggested probability of any restriction at all is 
$e^{-\frac{5^2}{2}}\simeq e^{-12.5}\simeq 10^{-\frac{12.5}{2.3}}
\simeq 10^{-5.4}\simeq 4\times 10^{-6}$.

To obtain a good answer to question 2) on the 
order-of-magnitude of the strength of the effect we must let the a priori 
probability of a drawing card giving a certain degree of restriction vary 
with the degree of restriction. Thus, a milder restriction is made, a priori, 
to be much more likely 
than a more severe restriction. We can basically 
assume that {\em we will not draw a restriction} (card combination) 
appreciably stronger than that required to demonstrate our effect. 
Thus, we can assume 
that the restriction drawn will be of the 
order of the magnitude of the strength of the operating machine, $\chi$, which is 
the maximum allowed before our effect stops it. 
Thus if we arrange the probabilities in this way, we may claim 
that the restriction resulting from the drawn random number represent the  
strength of the effect. Mathematically, such an arrangement 
means that we choose the a priori probability for the restriction value 
of $\chi$, $\chi_{\text{restriction}}$, to be a power law: 
\begin{align}
	P_{\text{a priori}} & \left(\left[\chi_{\text{restriction}}-\frac{1}{2}
	d\chi_{\text{restriction}},\chi_{\text{restriction}}+\frac{1}{2}
	d\chi_{\text{restriction}}\right]\right) \nonumber \\
	& =p\left(\chi_{\text{restriction}}\right)d\chi_{\text{restriction}}
~,
\end{align}
where
\begin{align}
	p\left(\chi_{\text{restriction}}\right)
	=K\chi_{\text{restriction}}^{\alpha}~, 
\end{align}
and $K$ is a normalization constant. Here a larger value of 
$\alpha >0$ should be chosen for a sharper measurement of the strength of the effect. 
If we only require a crude order of magnitude, we can simply take $\alpha 
\simeq 1$.

Note that having a large $\alpha$ means that 
very severe restrictions become relatively very unlikely. Thus, a large 
$\alpha$ is optimal for ensuring minimal harm to the operation of the machine.

We shall also assume that since the Tevatron seems not to be disturbed 
we do not have to include more severe restrictions on the LHC than 
those that would
force it to operate as a Tevatron.

Concerning question 3) as to which features of the operation, 
e.g., luminosity and center-of-mass beam energy, are the most important for our 
effect, we answer this question by letting 
different random numbers -- the drawings -- result in different {\em types of 
restrictions}. 
That is to say, the different drawings represent different 
combinations of restrictions on the beam energy and luminosity. Presumably it would 
be wise to make as many variations in the restriction patterns 
as possible, because the more combinations of various parameters,
the more information about our effect one can obtain. 
If one draws a combination 
of cards that causes a restriction, then one has immediately verified our type 
of model or the existence of an effect from the future. 
In this case, any 
detail of the specific restriction combination obtained from
the drawing is no longer 
random but is an expression of the mysterious new effect just established 
by the same drawing.  The more details one can thus arrange to be readable 
from the card combination drawn, the more information one will obtain about the 
$S_I$-effect in the case of restrictions
that actually show up in spite of 
having been a priori arranged to do so with a probability of the order 
of 5 standard deviations from the mean. 
Thus, to obtain as much profitable
information as possible, there should be as many 
drawing combinations with as many different detailed restrictions as 
possible. One could easily make restrictions only for a limited number of 
years or one could restrict the number of Higgs particles produced according 
to some specific Monte Carlo program using, at that time, the best estimate for 
the mass of a Higgs particle. If one allows some irrelevant 
details to also result from the drawing, it is not a serious problem, since one 
will simply obtain a random answer concerning the irrelevant parameter. It would be 
much worse if our theory was correct and one missed the chance of extracting 
an important parameter, that could have been extracted from the drawing.

One should therefore also be careful when adjusting the relative a priori 
probabilities of the values for parameters one hopes to extract, so 
that one really extracts interesting information relative to theoretical 
expectations and does not simply obtain a certain result because one has 
adjusted the priori probability too much.

Concerning question 4), to determine the type of random number 
that can be most easily manipulated by 
our $S_I$-effect, we should extract information -- but 
unfortunately very little information we suspect -- to answer the question, by using 
several, or at least two, competing types of random numbers. One could, for instance, 
have one quantum mechanical random number generator and one card-drawing 
game. One could easily reduce the probability of a restriction in each of them
by a factor of 2 
so as to keep the total probability of obtaining a restriction at the initially 
prescribed level of 5 standard deviations. Then one 
should have two (or more) sources of random numbers, e.g., a genuine 
card-drawing game and a quantum random number generator, each with a very high 
probability that there will be no restrictions on the running 
of the LHC (for that type of random number) and only a tiny 
probability of some restriction (as already discussed with as many different 
ways of imposing a restriction 
as one can invent) of less than 5 standard deviations 
divided by the number of different types of random number,  
2 in our example.

After having drawn a restriction from one of the types of 
random numbers, one would at least know that this 
type of random number was accessible for manipulation by our 
$S_I$-effect. Such information could be of theoretical value because one 
can potentially imagine that various detailed models based on our type of effect 
from a future model may give various predictions as to through which type of 
random number the $S_I$-effect can express itself. If, say, a model only 
allowed the $S_I$-effect through classical effects of the initial state of the 
universe but quantum experiments gave fundamentally random 
or ``fortuitous" \cite{Bohr} results that not even $S_I$ could influence, then 
such a model would be falsified if 
$S_I$-effect produced the quantum number
led to restrictions on the LHC.

One could also imagine that more detailed calculations would 
determine whether 
the effect from the future had to manifest itself not too far back in time. In 
that case one could perhaps invent a type of card game with cards 
that had been shuffled 
many years in advance, and one only used the first six cards in such stack of cards.

If it was the type of random number that came from stack shuffled
years in advance that allowed the $S_I$-effect,
then any type of detailed theory in which the effects of the future go only a short 
time interval back in time could be falsified.

\subsection{Statistical cost estimate of experiment so far discussed}

~~~~Let us now, as a first overview as to how risky it would be to perform a random 
number experiment, consider the simple proposal above: 

The highest probability in the experiment is that no
restrictions are imposed
because we only propose restrictions with a probability of the 
order of $10^{-6}$. 
Even in the case of drawing a restriction, 
one then considers the distribution of, say, the beam energy 
restriction to be the $\alpha$th power of the beam energy. Here we think 
of $\alpha$ being $1$ or $2$. This leads to the average allowed 
beam energy being reduced by  
$\chi_{\text{LHC}}\cdot \langle (1-X)\rangle$, where $X$ denotes the fraction 
of allowed beam relative to the maximum beam. In other words we call the highest 
allowed beam according to the card-drawing game $X\chi_{\text{LHC}}$. 
Then the average reduction in the case with $10^{-6}$ probability that we get a
reduction relative to the maximum beam $\chi_{\text{LHC}}$ becomes 
\begin{align}
	\langle 1-X\rangle 
	& =\frac{\displaystyle \int_{\text{Tevatron}}^{\text{LHC max}}
	X^{+\alpha}(1-X)dX}{\displaystyle 
	\int_{\text{Tevatron}}^{\text{LHC max}}X^{+\alpha}dX} \cr
	& =\frac{\left(\frac{1}{-\alpha -1}-\frac{1}{-\alpha -2}\right)
	\chi_{\text{LHC}}}{\chi_{\text{LHC}}\frac{1}{-\alpha -1}} \cr
	& =\frac{-\alpha -2-(-\alpha -1)}{-\alpha -2} \cr
	& =\frac{1}{\alpha +2}.
\end{align}
For $\alpha =2$ we lose $\frac{1}{4}$ of the maximum beam due to the restriction.

With the cost of the LHC machine estimated at $2$ to $3$ billion Swiss francs,
the probability of a restriction $r$ being 
$10^{-6}$, and 
the expected loss of the beam being
$\frac{1}{4}$, the average cost 
of the card game experiment is of the order of $100$ Swiss francs. 
However, of course there is, a 
priori, a risk. One shall, however, not mind if the 
``bad luck of drawing a restriction card" occurs, because in reality it is 
fantastically good luck because one would have discovered a 
fantastic and at first unbelievable effect from the future!

\subsection{Attempts to further reduce the harm}

~~~~One can, of course, seek to further bias the rules of the game so 
as to assign the highest probabilities to the restrictions causing least harm. 
For instance, one could allow a relatively high probability for 
restrictions of the 
type in which one is only allowed to operate the machine for a short time at its 
highest energy.

\section{Competition determining the fate between different times}

~~~~To determine how our effect from the future 
functions let us consider
our imaginary Lagrangian model from a more theoretical viewpoint.

In the classical approximation of our model, we consider a first approximation 
such that 
\begin{enumerate}
\item[1)] the classical solution is determined alone by extremizing the 
{\em real} part of the action $S_R[path]$, i.e., 
\begin{align}
	\delta S_R[path\> cl. sol.]=0
~.
\end{align}
The reason for this is very simple. The real part $S_R$ determines 
the phase variation of the integral in the Feynman pathway 
\begin{align}
	\exp \left\{i\left(S_R+iS_I\right)\right\}
~,
\end{align}
and thus, it is only when $S_R$ varies slowly, i.e., when 
$\delta S_R\simeq 0$, that we do not have huge cancellation because the 
rapid sign variation (phase rotation) cancels the contribution out.

We may illustrate this by the following drawing. 
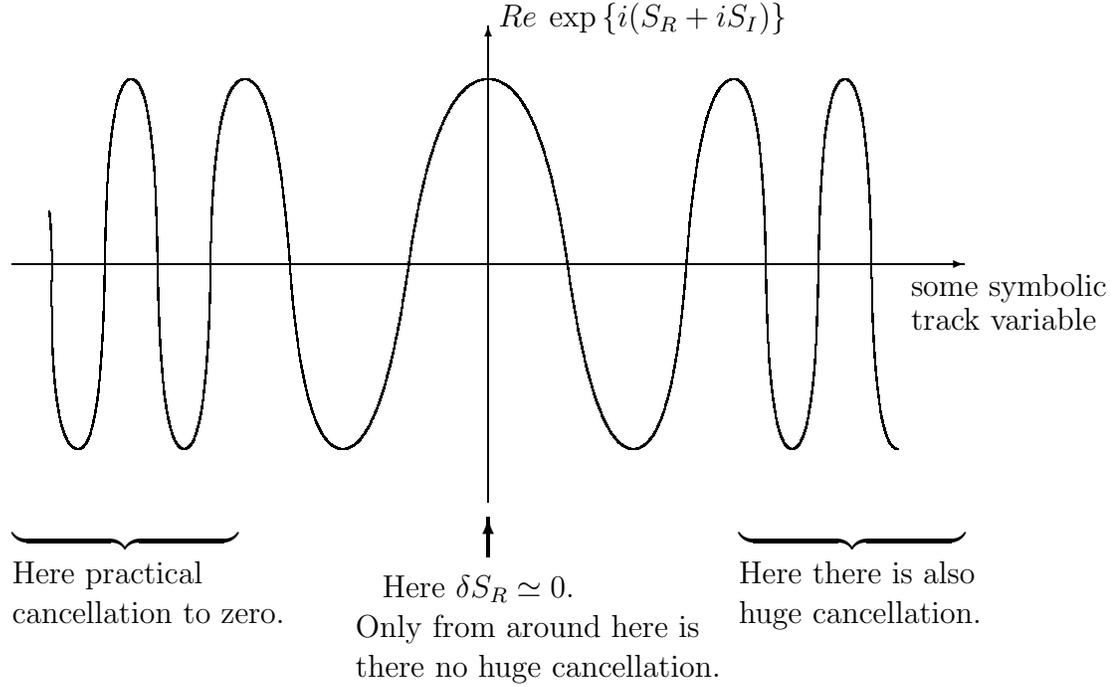
\begin{center}
\begin{picture}(380,200)
	\put(10,100){\vector(1,0){360}}
	\put(190,10){\vector(0,1){180}}
	\put(194,190){$Re\> \exp \left\{i(S_R+iS_I)\right\}$}
	\put(350,88){some symbolic}
	\put(350,75){track variable}
	\qbezier(190,170)(170,170)(160,100)
	\qbezier(160,100)(150,30)(135,30)
	\qbezier(135,30)(120,30)(115,100)
	\qbezier(115,100)(110,170)(98,170)
	\qbezier(98,170)(86,170)(85,100)
	\qbezier(85,100)(84,30)(75,30)
	\qbezier(75,30)(66,30)(65,100)
	\qbezier(65,100)(64,170)(55,170)
	\qbezier(55,170)(46,170)(45,100)
	\qbezier(45,100)(44,30)(35,30)
	\qbezier(35,30)(24,30)(25,100)
	\qbezier(25,100)(25,110)(24,120)
	\qbezier(190,170)(210,170)(220,100)
	\qbezier(220,100)(230,30)(245,30)
	\qbezier(245,30)(260,30)(265,100)
	\qbezier(265,100)(270,170)(283,170)
	\qbezier(283,170)(294,170)(295,100)
	\qbezier(295,100)(296,30)(305,30)
	\qbezier(305,30)(314,30)(315,100)
	\qbezier(315,100)(316,170)(325,170)
	\qbezier(325,170)(334,170)(335,100)
	\qbezier(325,170)(334,170)(335,100)
	\qbezier(335,100)(336,30)(345,30)
\end{picture}
\begin{picture}(380,70)
	\put(190,60){\thicklines \vector(0,1){15}}
	\put(150,45){Here $\delta S_R\simeq 0$.}
	\put(140,30){Only from around here is }
	\put(140,15){there no huge cancellation.}
	\put(10,70){$\underbrace{\text{\hspace{3cm}}}$}
	\put(10,50){Here practical}
	\put(10,35){cancellation to zero.}
	\put(285,70){$\underbrace{\text{\hspace{3cm}}}$}
	\put(285,50){Here there is also}
	\put(285,35){huge cancellation.}
\end{picture}
\end{center}

$S_I$ only results in a factor in the magnitude and leaves the phase of the 
integrand undisturbed.
\item[2)] The effect of $S_I$ has a total weight of $e^{-S_I[cl.sol.]}$ 
on the amplitude or Feynman path integral contribution, upon which one then inserts 
the solution to the classical equations 
of motion (i.e., to $\delta S_R=0$) 
for the path in the symbol $S_I[path]$.
This means then that the probability that the classical solution
``$cl.sol.$" exists is proportional to 
$e^{-2S_I[cl.sol.]}$. 
We have to square the amplitude to obtain the probability.

\indent The probability density of $e^{-2S_I[cl.sol.]}$  was referred to as  
$P[cl.sol.]$ in our early works in the present series and, unless one adds 
special assumptions about $S_I$, it behaves as a function of the path, i.e.,
there is only notational difference between $P[cl.sol.]$ in the early works and 
$e^{-2S_I[cl.sol.]}$, i.e., 
\begin{align}
	P[cl.sol.]=e^{-2S_I[cl.sol.]}.
\end{align}
\end{enumerate}

\subsection{The importance of competition between times on determining the fate 
of all times}

~~~~Here we want to stress a very important effect that reduces
the strength 
of the observable effect from the future in our model.
We call this effect
the competition between the different eras 
upon what shall happen in the universe.
We have noted that the probability of a certain classical solution to the 
equations of motion $\delta S_R=0$  -- the true track -- 
is given by $e^{-2S_I[cl.sol.]}$, where the imaginary action $S_I$ is 
{\em an integral over time} 
\begin{align}
	S_I=\int L_Idt.
\end{align}

The important point here is that selecting a certain solution to the 
equations of motion in one period of time via the equations of motion 
in principle determines the solution at all times, both earlier and later. 
This is basically ``determinism"; simply knowing the position and 
velocities of all the dynamical configurations of variables at one moment of 
time allows one, in principle, to integrate the equations of motion so as to 
obtain the solution at all times. This determinism may only be  
true in a principal or in an ideal way, since we know that, depending on the 
Lyapunov exponents, very small deviations between two solutions at one time can
become huge at a later time. Also, extrapolation backward 
in time may also have the same effect. Furthermore, it is known 
that this determinism is challenged by the measurement postulates of 
randomness in quantum mechanics.

Nevertheless, these is certainly a strong restriction as to what can be obtained from 
a solution at one moment of time if it has already been used to make a 
small $L_I$ at another time. The different regions in time are, so to speak, 
competing in the selection of the solution that gives the minimal contribution to 
$\int_{\text{a time region}}L_Idt$ in the different time regions. 
Here, we simply draw attention to the ``competitional" problem that the 
contribution to $S_I$ from $\int_{\text{Big Bang time}}L_Idt$ in Big Bang 
time does not usually make $\int_{\text{our times}}L_Idt$ the minimal value. 
Thus, a compromise must occur between the different time eras so as to 
minimize  
\begin{align}
	S_I=\int_{\text{all times}}L_Idt.
\end{align}

This means that even if one estimates a large effect of $L_I$ in one era, it may 
not be easy to use this effect to determine the minimal $L_I$ in our times 
or $\int_{\text{our times}}L_Idt$, because the enormously long time spans 
outside our own times will typically almost completely determine which 
solution to $\delta S_R=0$ will be selected that results in 
minimal $S_I$. Our own 
human lifetime only makes up an extremely small part of the 
$10^{10}$ years in which the universe has already existed. Thus,  
much stronger $S_I$-contributions are needed to have any effect than 
would be required with a universe existing only for human-scale time.

In other words, practically everything about the solution obtained by minimizing 
$S_I$ is determined by contributions from time 
intervals very far from the interval in which we know some history and 
have some memory 
of its significances. This means that we should observe extremely little 
effect from the future in practice. We would not be able to recognize much 
of any effect because most of the future as well as most of the past 
is so remote that we 
know exceedingly little about it.

We might recognize an effect from the future if some accelerator that is 
already planned is then stopped by Congress. However, 
if the accelerator is to be built in $10^{10}$ years, we would most likely 
not know about the plans and be unable to recognize the effect from the 
future that causes its closure in $10^{10}$ years from now.

We would only see such prearrangements as purely random and not as 
prearrangements. We can conclude that prearrangement is difficult to recognize unless 
you have knowledge of the plans that shall be accepted or rejected.

\subsection{The rough mathematical picture}

~~~~To get an idea of the significance of the competition between various time 
intervals on determining what is selected to be true solution of the classical 
equations of motion, we give a very rough description of the mathematics 
involved in
minimizing $S_I$ and in searching for a likely type of solution when we 
have the probability density $e^{-2S_I}$ over phase space. One should bear in 
mind that the set of all classical solutions are in one-to-one correspondence 
with phase space points when a solution is given by integrating up 
-backward and forward- from a certain standard moment $t_0$.

We should take $S_I$ to be an integral or a sum 
over a large number of small time intervals $\sum_i\int_{I_i}L_Idt$. Each of these 
small contributions $\int_{I_i}L_Idt$ may be taken as a random function 
written as a Fourier series over phase space in the very rough approximation 
for the first orientation. We even assume for the first orientation 
that we have a random form of $L_I(t_i)$ or, approximately equivalently, 
$\int_{I_i}L_Idt$ remains of the same form with the same probability of having 
different values after being transformed to the standard moment $t_0$ by using 
the canonical transformations associated with the Hamiltonian derived from the 
real part of the action $S_R$. That is to say, in the phase space 
variables $(\vec{q}_0,\vec{p}_0)$ at time $t_0$, each of the contributions 
$\int_{I_i}L_Idt$ is a stochastic variable function over phase space that can 
be expressed as 
\begin{align}
	\int_{I_i}L_Idt=\sum c(\vec{k}^{(q)},\vec{k}^{(p)})\cdot 
	e^{i\vec{k}^{(q)}\cdot \vec{q}+i\vec{k}^{(p)}\cdot \vec{p}}, 
\end{align}
where we require 
\begin{align}
	c(-\vec{k}^{(q)},-\vec{k}^{(p)})^{\ast}=c(\vec{k}^{(q)},\vec{k}^{(p)})
.\end{align}
We take the real and imaginary parts of the $c(\vec{k}^{(q)},\vec{k}^{(p)})$ 
to have a Gaussian distribution with the spread 
\begin{align}
	& \langle \left\{Re\> c(\vec{k}^{(q)},\vec{k}^{(p)})\right\}^2\rangle 
	=\sigma_r(\vec{k}^{(q)},\vec{k}^{(p)}), \nonumber \\
	& \langle \left\{Im\> c(\vec{k}^{(q)},\vec{k}^{(p)})\right\}^2\rangle 
	=\sigma_i(\vec{k}^{(q)},\vec{k}^{(p)}).
\end{align}
Since it is a rough model, we shall not go into details of how to 
choose $\sigma_r$ and $\sigma_i$, but imagine that we have a cutoff
that effectively separates large 
$\vec{k}^{(p)}$ and $\vec{k}^{(q)}$ regions. Also 
we would like to roughly take each $\int_{I_i}L_Idt$ to be 
periodic in the phase space variables $\vec{q}$ and $\vec{p}$ so that we 
effectively use a Fourier series. The period is a cutoff in phase space 
$\Lambda_{phs}$, and the cutoff in $\vec{k}^{(q)}$ and $\vec{k}^{(p)}$, 
say $\Lambda_k$, separates the 
rapid variations of $\int_{I_i}L_Idt$ over the phase 
space. There is, thus, effectively a number of independent phase space points 
$\left(\Lambda_k/\Lambda_{phs}\right)^N$ 
in which $\int_{I_i}L_Idt$ can take its values. 
Here $N$ is the number of degrees of freedom. The statistical 
distribution for one of the $\int_{I_i}L_Idt$ in one of these effective 
phase space points is assumed to be Gaussian with a mean square deviation of  
\begin{align}
	\sigma_r=\sum_{\vec{k}^{(p)},\vec{k}^{(q)}}
	\sigma_r(\vec{k}^{(q)},\vec{k}^{(p)})\simeq 
	\left(\frac{\Lambda_k}{\Lambda_{phs}}\right)^N\cdot \sigma_r
~,\end{align}
where $\sigma_r$ is a typical value for 
$\sigma_r(\vec{k}^{(q)},\vec{k}^{(p)})$. The contribution from the imaginary 
part is of the same order, and we ignore a factor of $2$ here.

When we search for $S_I=\sum_i^{n_{step}}\int_{I_i}L_{I_i}dt$, we again have a 
Gaussian distribution since each $L_{I_i}$ has,
by assumption, independent Gaussians. 
But if  
there are $n_{step}$ time steps of type $I_i$ then the distribution of 
$S_I$ becomes broader than that of $\int_{I_i}L_Idt$ by a factor of 
$\sqrt{n_{step}}$. That is to say, $S_I$ is of the order 
$\sqrt{n_{step}}\cdot \sigma_r$.

The mathematical picture, we have constructed is summarized by the following 
two points: 
\begin{enumerate}
\item[1)] There are $\left(\Lambda_k/\Lambda_{phs}\right)^N$ classical path solutions, 
or equivalently $\left(\Lambda_k/\Lambda_{phs}\right)^N$ possible ways that 
the universe could have started and subsequently developed.
\item[2)] $S_I$ for each of these developments has a Gaussian 
distribution with a mean square deviation of $n_{step}\sigma_r$.
\end{enumerate}

Our model postulates that the probability of the realization of a classical solution 
is weighted by $P=e^{-2S_I}$.

This extra effect of our model converts the distribution of what from a Gaussian of 
the form 
\begin{align}
	\exp \left(-\frac{S_I^2}{2n_{step}\cdot \sigma_r}\right)
\end{align}
into a distribution of the form
\begin{align}
	\exp \left(-\frac{S_I^2}{2n_{step}\cdot \sigma_r}-2S_I\right).
	\label{eq21}
\end{align}
Equation (\ref{eq21}) implies that the realistic or most likely $S_I$-value for the 
chosen development of the universe should be realized by maximizing the 
exponent 
\begin{align}
	-\frac{S_I^2}{2n_{step}\cdot \sigma_r}-2S_I
\end{align}
in this probability distribution.

The maximum occurs when 
\begin{align}
	S_I\simeq n_{step}\sigma .
\end{align}
The actual development of the universe will not have exactly this value 
$n_{step}\sigma$ for $S_I$, but a value typically deviating by the 
order $\sqrt{n_{step}\sigma}$.

Now let us imagine that in our time, say, in one of the 
intervals $I_i$, we look for  
a special occurrence that may give an extra 
contribution $\Delta S_{I\> extra}$ to $S_I$. It is easy to see 
that compared with the probability without this contribution, the probability 
with the $\Delta S_{I\> extra}$ contribution should be 
$e^{-2\Delta S_{I\> extra}}$ times high. However, the question is whether 
we would realistically notice this effect.

To detect our $S_I$-effect we might carry out a card-drawing experiment by 
turning a card, which if black we let the experiment giving 
$\Delta S_{I\> extra}$ be performed, 
and if red we do not perform it.
We first imagine, for the sake of 
argument, that the extra $L_I$ contribution is switched off by not there at all. 
Then, for symmetry 
reasons, the probabilities of red and black should both be $\frac{1}{2}$.

Thus, it would appear that the result of black or red would be 
dominantly determined from what happens in other time intervals rather than in 
``our time", in the sense that the contribution to the fluctuation in $S_I$ 
from times other than ours would be 
\begin{align}
	\Delta S_{I\> fluctuation}\Big|_{\text{from other times}}
	=\sqrt{(n_{step}-1)\sigma}.
\end{align}

This effect of the contributions from these other times, which we cannot treat 
scientifically or understand or know anything significant about, will give 
an $S_I$-contribution of the order 
$\sqrt{(n_{step}-1)\sigma}\sim \sqrt{n_{step}\sigma}$, to which the extra 
contribution $\Delta S_{I\> extra}$ has to be compared when it is switched on.

Let us first estimate what we would be an ordinary value of  
$\Delta S_{I\> extra}$ relative to $\sqrt{\sigma}$. Since we not only live in 
a short accessible time but also in a small accessible spatial region, we 
should take an ordinary order of magnitude for $\Delta S_{I\> extra}$ of 
$\sqrt{\sigma \cdot \frac{V_{acc}}{V_{univ}}}$, where $V_{acc}$ is the 
part of the universe controllable by our card game and $V_{univ}$ is the 
total effective volume of the universe.

Thus, the ``ordinary" value for $\Delta S_{I\> extra}$ is 
\begin{align}
	\Delta S_{I\> extra}\Big|_{\text{ordinary}}\sim 
	\sqrt{\sigma \cdot \frac{V_{acc}}{V_{univ}}}
~,\end{align}
which is to be compared with 
\begin{align}
	\Delta S_{I\> fluctuation}\sim \sqrt{n_{step}\sigma}~.
\end{align}
This gives 
\begin{align}
	\frac{\Delta S_{I\> extra}\Big|_{\text{ordinary}}}
	{\Delta S_{I\> fluctuation}}\simeq 
	\sqrt{\frac{V_{acc}}{n_{step}V_{univ}}}~,
\end{align}
which means the square root of the universe accessible by the card game part of space 
time relative to the full space time of the universe.

If we use a weak scale to give us the region in space time in which a Higgs 
particle contributes 
$l_W\sim \frac{1}{100\text{ GeV}}\sim 2\times 10^{-3}\text{ fm}
\sim 10^{-26}\text{ s}$ while the extension of the  reachable universe 
and its lifetime is taken to be $10^{17}$ s then 
\begin{align}
	\frac{V_{acc}}{n_{step}V_{univ}}\sim 
	\left(\frac{10^{-26}\text{ s}}{10^{17}\text{ s}}\right)^4
	=10^{-172}.
\end{align}
This would give us $10^{-86}$ as the quantity that must  be 
compensated by 
having an extraordinary size of $L_I$
to obtain any recognizable effect. Now if, as is quite likely, 
the hierarchy-problem-related  fine tuning of the square of the 
Higgs mass should only be for the 
real part $m_{HR}^2$ of the square of the mass, 
while the imaginary part $m_{HI}^2$ 
of the $|\phi_H|^2$-coefficient is of the Planck scale order of 
magnitude, $m_{HI}^2\sim (10^{19}\text{ GeV})^2$, then the ratio of 
$\Delta S_{I_{extra}}$ relative to $\Delta S_{I_{ordinary}}$
from the $m_{HI}^2$-term would be expected to be of the 
order of 
\begin{align}
	\frac{m_{HI}^2}{m_{HR}^2}\sim 10^{34}
\end{align}
times bigger than ``ordinary" $S_I$-contribution. This would not be enough to 
compensate the $10^{-172}$, but the latter might be very many orders of 
magnitude wrong for several reasons, as we shall now discuss in the next subsection.

\subsection{What value to take for an effective $\frac{V_{acc}}{n_{step}V_{univ}}$?}

~~~~One could say: 
\begin{enumerate}
\item[1)] The card game result is mainly connected with the earth 
as far as its development and dependence is concerned.
Thus we should reduce
the 
effective universe size $V_{univ}$ to be that of the earth, i.e., a 
length scale of $10^7\text{ m}\sim 3\times 10^{-2}\text{ s}$ rather than the 
$10^{17}\text{ s}$ in the above estimation.
\item[2)] If we compare the situation on earth events only really occur 
in the atoms, and if something happens at a weak scale when 
Higgs particles are present it may seem reasonable that each Higgs particle 
has as many degrees of freedom as an atom and should be assigned in our 
estimate space having an as atomic size, i.e., $\sim 10^{-10}\text{ m}
\sim \frac{1}{3}\times 10^{-18}\text{ s}$.

From only these two corrections we would obtain 
\begin{align}
	\frac{V_{acc}}{V_{univ}}\Bigg|_{\text{eff}}\sim 
	\left(\frac{\frac{1}{3}\times 10^{-18}\text{ s}}
	{3\times 10^{-2}\text{ s}}\right)^3\simeq 10^{-45}
.\end{align}
\item[3)] We might also have to count the lifetime of the 
Higgs particle as closer to 
$\frac{1}{\text{MeV}}$ than $\frac{1}{100\text{ GeV}}$ meaning that 
$n_{step}$ would decrease from 
$n_{step}\sim \frac{10^{17}\text{ s}}{10^{-26}\text{ s}}\simeq 10^{43}$ 
to $n_{step}\sim \frac{10^{17}\text{ s}}{10^{-21}\text{ s}}\simeq 10^{38}$. 
This would increase the square root to 
$\sqrt{\frac{V_{acc}}{n_{step}V_{univ}}}
\simeq \sqrt{\frac{10^{-45}}{10^{38}}}
\simeq \sqrt{10^{-83}}\simeq 10^{-41.5}$. Then we would only lack a factor of 
$\frac{10^{41.5}}{10^{34}}=10^{7.5}$, which may be able to compete in 
producing a significant value of $\Delta S_{I\> extra}$ 
due to the existence of many Higgs particles.
\item[4)] If, for instance, 
we could replace the whole lifetime of the universe 
by some inverse Lyapunov exponent for political activities, 
i.e., the time in which exceedingly small 
effects develop into politically important decisions,
say a few years, then we could effectively reduce $n_{step}$ by a factor 
$10^9$ and we would need $10^{4.5}$ times less compensation.

This would mean that we might only need to produce $10^3$ Higgs particles 
in an accelerator for it to be enough that the SSC would be canceled by 
our model.

In the light of the huge uncertainties even in the logarithm of these 
estimates and the closeness to the achievements of the LHC of our relevant scale, 
it is clear that a much better estimation to the extent that such an estimate
is possible is called for.
\end{enumerate}

\subsection{Baryon destruction}

~~~~Let us bring attention here to an effect in our model that will potentially be 
much more important than the Higgs particle production in the SSC: baryon destruction. 
Since the quarks in the baryons couple to the Higgs particle field, which 
decreases it numerically to close to that of the quark or baryon, they function 
as a tiny negative number of Higgs particles. However, in contrast to the Higgs 
particle itself, the baryon effectively lives eternally. Thus, if one produces an 
accelerator that has sufficiently high energy that it can violate the baryon number, then 
it may affect the Higgs field more than genuine Higgs particles. Indeed, the 
$|\phi_H|^2$-charge by $d$ quark may typically be of the order of 
$|g_d|^2$ times that of a genuine Higgs charge. However, since the baryon lives 
eternally, we gain a lifetime factor of 
$\frac{10^7\text{ s}}{10^{-21}\text{ s}}=10^{38}$, which hugely 
overcompensates for $|g_d|^2\sim (10^{-5})^2=10^{-10}$. 
Thus, the destruction of a single 
baryon should be about as $S_I$-significant as $10^{28}$ Higgs particles. 
This estimate would result in the borderline significance  of the Higgs 
particle being converted into an absolute necessity for the SSC to be 
canceled.

Now we might even ask whether our first estimate 
\begin{align}
	\sqrt{\frac{V_{acc}}{n_{step}V_{univ}}}\Bigg|_{\text{first}}\simeq 
	10^{-86}
\end{align}
would allow there to be $S_I$-effects resulting from baryon destruction. Because 
of the extremely long baryon lifetime compared with the Higgs particle, 
the effect of $\Delta S_{I\> extra}$ increased by a factor of $10^{28}$ 
upon the destruction of a baryon. 
This would convert $m_{HI}^2/m_{HR}^2\sim 10^{34}$ into a 
factor greater than the ``ordinary" one, $10^{34+28}=10^{62}$. Even that would not be 
sufficient to compensate for $10^{-86}$.

However, even one of the above corrections results in a change from the weak size 
to the atomic size, thereby increasing $V_{acc}$ by a factor of $(10^7)^3$, and 
thus increasing $\sqrt{\frac{V_{acc}}{n_{step}V_{univ}}}$ by 
a factor of $10^{10.5}$ or perhaps more 
correctly, using the distance between the atoms in the universe, resulting in yet 
another increase by a similar factor $\sqrt{(10^8)^3}=10^{12}$. 
Indeed, we would then have 
$\sqrt{\frac{V_{acc}}{n_{step}V_{univ}}}\sim 10^{-65}$, so that the $10^{62}$ 
could cope if the the SSC had destroyed only $10^3$ baryons. However,  
if the baryon destruction resulted in the cancellation of the SSC, then the LHC is not 
in danger because there will presumably be no baryon destruction at the LHC.
Presumably there would not even have been in the SSC.

\section{Conclusion}

~~~~We first reviewed our model with an imaginary action 
to be inserted into the  
Feynman pathway integral. It seems a bit artificial to assume that the action $S$ 
in the integrand $e^{\frac{i}{\hbar}S}$ should be wholly real when 
the integrand itself is clearly complex. We claim that such an imaginary 
part $S_I[path]$ in the action $S=S_R+iS_I$ does not influence the classical 
equations of motion $\delta S_R=0$, but rather manifests itself by determining
the initial 
conditions for the development of the universe. Indeed, the various classical 
solutions that contribute to the Feynman pathway integral are weighted 
by an extra factor 
$e^{-S_I[path]}$, which leads to a probability weight of $P=e^{-2S_I[path]}$.

The main discussion in the present article was on the development of an earlier 
proposal \cite{search} for how to search for the effects on the determination of 
initial conditions of such an imaginary action term $S_I[path]$. 
From this viewpoint, 
the most remarkable fact is that this imaginary part $S_I$, in analogy 
with the real part $S_R$, is given as an integral 
\begin{align}
	S_I=\int L_Idt
\end{align}
over all times and thus depends on the fields or dynamical variables not only 
in the past but at all times. Thereby, the discussion became focused on searching for 
effects from the initial conditions, which have been adjusted so as to take into 
account what should happen or should not happen at a much later time.

Because of the very high probability that, in contrast to the real part $m_{HR}^2$ 
of the coefficient $m_{H}^2=m_{HR}^2+im_{HI}^2$ of the expression $|\phi_H|^2$ 
in the Higgs field $\phi_H(x)$ part of the Lagrangian density 
$\CL(x)=\CL_R(x)+i\CL_I(x)$, the imaginary part $m_{HI}^2$ was not fine tuned to be 
exceedingly small compared with the fundamental scale. It was suggested that 
$m_{HI}^2$ is likely to be huge compared with $m_{HR}^2$. We refer here to the 
problem behind the so-called hierarchy problem associated with the fact that the 
weak-energy scale given by $m_{HR}^2$ is very small compared with, say, the 
Planck scale. Since the reason for this fine tuning of $m_{HR}^2$ to a small value 
is still unknown, it may be equally likely that the mechanism 
for this fine tuning would also tune $m_{HI}^2$ to a small value or leave it at 
the Planck scale. It is therefore very likely that there is an imaginary action 
for which the ratio between the corresponding coefficients in the imaginary part $S_I$ 
and the real part $S_R$ would be unusually large in the case of the Higgs mass 
square say 
\begin{align}
	\frac{m_{HI}^2}{m_{HR}^2}\sim 10^{34}.
\end{align}

This possibility makes it likely that particularly large effects of 
$S_I$ can be found, 
and thus, cases of the future influencing even the initial conditions, and thus 
the past may occur when the Higgs mass square term is involved. In almost all 
investigations of 
the Standard Model so far, the Higgs mass square term 
$(m_{HR}^2+im_{HI}^2)|\phi_H|^2$ is only involved via the Higgs field vacuum 
expectation value $\langle \phi_H\rangle$, which is determined only from the 
real part $m_{HR}^2$. Thus, we may have to wait for genuine Higgs-particle-producing machines to search for the effects of the huge 
expected imaginary part $m_{HI}^2$ of the Higgs mass square. 
Alternatively, we would have 
to search for the effects of previously observed particles, such as quarks or 
leptons, on the Higgs field, by a back reaction which  
give a contribution proportional to $m_{HI}^2$ to $S_I$. 

One such effect 
could be caused by an accelerator able to violate the conservation of the 
baryon number \cite{baryonviolation} and presumably destroy more baryons than 
it creates. Then, the width $|g_d|^2$ -- the quark Yukawa coupling squared -- 
which is proportional to the 
suppression of the Higgs field around the baryon or the quark 
would be lost for the rest of the existence of the universe. This effect 
of having a baryon being destroyed forever while a Higgs particle 
has only a short lifetime overcompensates for the Yukawa coupling 
suppression so that the effect of a baryon being destroyed on $S_I$ is presumably 
much bigger -- by say $10^{28}$ -- than that of the creation of a genuine Higgs particle. 
Nevertheless, our estimates of these effects are at the moment so approximate that it 
is uncertain whether the $S_I$-effect would be sufficient for 
preventing Higgs production; thus,
we predict possibly that the initial state would have been 
organized somehow so that a large Higgs-particle-producing 
machine such as the LHC should 
somehow be prearranged so as not to come into existence.

Such an effect would, of course, be 
even stronger for the the terminated SSC machine in Texas 
since it would not only have produced more Higgs particles but also perhaps have 
destroyed some baryons.

From the LHC-threatening perspective, the main point of the present 
article is that the LHC should really not be allowed 
to operate at full intensity or beam energy
by these effects on the initial state due to $S_I$. 
In order to obtain as much knowledge and as little loss as possible out of this 
otherwise problematic event.
We then propose our card or random number experiment.

Our main proposal was to perform a quantum random number or card-drawing 
experiment, with both ``old" and ``new" random numbers, meaning 
that the random numbers are created at longer or shorter times before the 
LHC is switched on, and then let the value of this random number determine the
restrictions on the running of the LHC.

It should be stressed that this whole process of closing random numbers to decide 
the fate of  the
LHC should be arranged so that it is by far the most likely that 
no restrictions are imposed at all. Only if there is some mysterious effect 
such as the $S_I$-effect in our model, which might have prearranged the initial state 
so as to prevent the LHC from operating, should there be any significant 
chance of obtaining anything apart from ``everything is allowed for the LHC". 
In this way, if any restriction was indeed drawn by the 
card or by the quantum random number generator then this would, a priori, be so 
unlikely that such a drawing would immediately justify our type of model.  
Such a result would be so miraculous that it would require a new set of physical laws.

Thus, we stress that if such a restriction is 
drawn, we should arrange matters so that
the exact value of the drawing tells us as much as 
possible about the details of the theory on the effect from the future,  which 
is justified merely by the selection 
of a restriction-requiring card.

This extraction of extra information from the drawing should have three features: 
\begin{enumerate}
\item[1)] One should use different types of random numbers such as a) cards shuffled 
recently, b) cards shuffled long ago, c) quantum random numbers made immediately 
before the decision, and d) quantum random numbers made in advance. 
Then one can determine which type of random number -- old/recent, 
card-drawing game/quantum -- is the easiest for the $S_I$-effect to 
manipulate so as to carry out the desired task of stopping or restricting the LHC.
\item[2)] One would like to know which parameters of the machine are 
important in terms of the $S_I$-effect. We should arrange the experiment so that, 
for all the different types of random numbers, 
there are appreciably higher a priori probabilities (i.e., higher 
numbers of card combinations) of mild restrictions 
than of strong restrictions. By designing the game in this way,
we can learn from the result which restriction is really needed
for causing any effect backward in time.
The trial is of course also economical in the sense that we would thereby only obtain 
a result giving the minimal restriction needed
to verify our theory; for example, this
may be that only high 
luminosity combined with the highest energy in the beams would be forbidden by 
the card, but many of other combinations of operational parameters would be 
allowed. This would 
then cause minimum disruption to 
the program at the LHC.
\item[3)] The most important result from the card-drawing experiment 
or random number selection if our model turned out to be 
correct may be that we would 
obtain a controllable estimate of its reliability. Of course, one would become 
convinced of a model of our type in which Higgs particles or baryon 
destruction affect the past if something happened so 
that the LHC was prevented from operating.
However, if this was not due to a controlled 
card-drawing or random number game, one could always claim
that the cooling system was defective, or that the failure was due to the
political circumstances of some physicists
or because of a war or an 
earthquake.
However, the $S_I$-effect would always have to 
use some natural effect to cause the effective restriction. Thus, there will 
always be alternative reasons that may account for the failure of the plant. 
The main point, of course, is that if our model is not 
true then by the far the most likely outcome 
is that the LHC will simply start operating next year.
Thus, if something happens to prevent the LHC from operation, 
then we should believe 
our theory, but to obtain a more easily estimable basis for the degree of 
belief if a failure or restriction occurred, we should carry out a fully 
controlled experiment involving 
random numbers, and if we had assigned an extremely low 
probability to the restriction that occurred, we would obtain valuable information
about the $S_I$-effect.

~~~~Of course, there is the ``danger" that by making the probability of 
an experimental restriction 
extremely low, it becomes more likely that if our theory was 
correct that, in spite of the selection, something else 
(the cooling system not working, an earthquake, etc.) would stop the machine.

~~~~In the present article we have also derived a very rough 
estimation of the amount of extra $S_I$, called $\Delta S_{I\> extra}$, 
that is needed to be significant enough to produce observable effects. So far, these 
estimates are so approximate that we cannot say that, 
even if our model was in principle 
correct, it would have sufficiently strong effects associated with the Higgs 
particle that it would actually lead to a closure or restriction of the LHC. 
We concluded from the approximation 
that the effect of the destruction of baryons, which has been 
speculated \cite{baryonviolation} would have occurred
at the SSC, would be much stronger than 
the effect of Higgs particles. 
However, in our first estimate, uncertainties were so 
high that we cannot even claim that if our theory was in principle 
correct, an accelerator causing baryon destruction would definitely have to be 
closed (by some sort of ``miraculous" effect).

~~~~We think that we may be able to produce a somewhat better estimate, but Lyapunov exponents 
in the real world, which includes
political decisions, may be difficult to estimate.
Of course, the real uncertainty is whether our model is true, even in principle. 
To establish the truth of such models,
successful cosmological predictions are hoped for. 
We are on the way to obtaining some predictions 
concerning the neutron lifetime by considering the order of the capture time 
in Big Bang nuclear synthesis, and the ratio of dark energy density to full 
energy density, which is estimated
to be around $2/3$ to $3/4$.
In fact, predictions are being made for essentially  
the whole process of cosmological development except for the first inflation itself.
\end{enumerate}

\section*{Acknowledgments}

We acknowledge the Niels Bohr Institute, Yukawa Institute for Theoretical 
Physics, Kyoto University and CERN Theory Group for the hospitality they extended 
to the authors. This work is supported by Grants-in Aid for Scientific 
Research on Priority Areas, 763 ``Dynamics of Strings and 
Fields'', from the Ministry of Education, Culture, Sports, Science and 
Technology, Japan. We also acknowledge discussions with colleagues, especially 
John Renner Hansen, on the SSC.           


\end{document}